# Predictive wavelength tailoring of uniform GaSb-based quantum dots for emission at 1.55 μm

*Markus Peil, Maja Wasiluk, Ziemowit Olinkiewicz, Tymon Przychodni, Robert Matysiak, Teemu Taskinen, Joona Salonen, Abhiroop Chellu, Metin Patli, Joonas Hilska, Anna Musiał, Michał Gawełczyk, Mircea Guina and Teemu Hakkarainen**

M. Peil, T. Taskinen, J. Salonen, A. Chellu, M. Patli, J. Hilska, M. Guina, T. Hakkarainen
Optoelectronics Research Centre, Tampere University, Tampere, Finland
E-mail: teemu.hakkarainen@tuni.fi

M. Wasiluk, A. Musiał
Department of Experimental Physics, Wrocław University of Science and Technology, Wrocław, Poland
Z. Olinkiewicz, T. Przychodni, R. Matysiak, M. Gawełczyk
Institute of Theoretical Physics, Wrocław University of Science and Technology, Wrocław, Poland

Funding: FiGAnti project in QuantERA II EU Program (Grant Agreement No. 101017733), Strategic Research Council of Finland (Decision No. 361293), National Science Centre Poland (Project 2023/05/Y/ST3/00125), Strategic Research Council of Finland via "CryoLight" project (Decision No. 357351), Flagship Program PREIN (Decision No. 368650)

((Markus Peil and Maja Wasiluk contributed equally to this work.))

**ABSTRACT.** A detailed study of emission wavelength tailoring of GaSb-based QDs formed by InGaSb-filling of droplet-etched nanoholes in AlGaSb is presented. The study shows that the emission wavelength can be modified from 1.48 µm to the center of the telecom C-band at 1.55 µm by independently varying the QD composition and size. More specifically, the optical transition energy shifts linearly as a function of In-content of the QD material at a rate of −4.4 meV/In%, and with the number of monolayers (ML) of material used for filling the nanoholes, at −2.0 meV/ML. These experimentally observed energy shifts are well predicted by simulations yielding rates of −4.3 meV/%In and −2.1 meV/ML, respectively. For the simulation, a uniform In composition, low intermixing, and microscopically measured QD geometry is considered. Additionally, excellent ensemble QD uniformity, with unprecedented inhomogeneous broadening well-below 7 meV across all samples is demonstrated. Finally, photoluminescence of single-QDs reveals narrow excitonic emission lines of 13.8±6.7 µeV and low fine-structure splitting values reaching <10 µeV. These results identify GaSb-based LDE QDs as a tunable telecom platform for scaling quantum-photonic applications over long-haul optical fiber networks.

## INTRODUCTION

Quantum communication has gained significant interest in the growing digital information era, where secure data transfer and communication are paramount. The ability of photons to transfer coherent superposition of quantum states at the speed of light is exploited for secure quantum communication[1] and for connecting remote quantum processors in distributed quantum computing networks[2]. To this end, operating at telecom wavelengths would enable efficient photon transfer through existing optical fiber infrastructure, which have minimal propagation losses at 1.55 µm[3]. Moreover, a slightly longer emission wavelength is relevant when implementing free-space optical links utilizing the ~1.6 µm atmospheric transmission window[4]. For generation of single photons and entangled photon pairs at these wavelengths, which are key building blocks for quantum communication, the current technology relies on spontaneous parametric down-conversion processes[5]. However, this method still faces significant challenges in achieving high state purity with high efficiency, thus limiting practicable brightness while simultaneously being fundamentally non-deterministic[5,6]. To address this limitation, epitaxial semiconductor quantum dots (QDs) have emerged as a leading source of single and entangled photons with high emission rate[7], minimal fine-structure splitting (FSS)[8], low multi-photon emission probabilities[9] and high entanglement fidelities[10–13]. Moreover, this approach would leverage established III-V semiconductor process technology, enabling fabrication scaling and interfacing with photonic integrated circuits[14–16]. However, the search for telecom-emitting QD material systems with sufficiently low density and high symmetry necessary for these applications has proven to be a real challenge. To circumvent this issue, some approaches have relied on frequency down conversion of shorter wavelength photons from GaAs and InGaAs QDs[17–19] to the telecom wavelengths, which has resulted in successful demonstrations of quantum-secure communication[20,21]. Still, QD sources emitting directly at telecom wavelengths would be a more straightforward solution with the benefits of reducing overall system costs, complexity, loss, and size with a simultaneous increase in source efficiency. Therefore, the search for the ideal telecom QD source has received increasing attention over the recent years[22–27] and has now shown notable achievements in several key figures of merit. For example, Stranski-Krastanow (SK) InAs/InP QDs have displayed low multi-photon emission probability with second-order correlation function value at zero time delay of $g^{(2)}(0) = (4.4 \pm 0.2) \times 10^{-4}$,[28] as well as high Hong-Ou-Mandel visibilities of 0.98[14] and 0.917[29] for structures incorporating an InGaAlAs quaternary barrier. Furthermore, SK InAs/InGaAs/GaAs QDs[30] have been used to demonstrate intercity BB84 quantum key distribution with a high 10.9 kbits/s key rate and a low quantum bit error ratio of 0.65%[31]. Despite these demonstrations, such QD material platforms face unique challenges that are detrimental when scaling in practical systems is considered. SK InAs QDs require either metamorphic buffers[30] or strain-reducing InGaAs layers on GaAs[32], or alternatively growth on InP[33,34], which results generally in broader size distribution and larger FSS than in state-of-the-art GaAs/AlGaAs QDs emitting at shorter wavelengths[8,35,36]. Additionally, while SK QDs with sufficiently low densities for single-QD devices have been achieved, they are extremely sensitive to growth parameters, impairing reproducibility and wafer-scale uniformity[37,38]. Another alternative is droplet epitaxy (DE) based QDs, which have shown emission at telecom wavelengths with improvements to the FSS compared to SK QDs[39,40], but challenges remain as the optical quality suffers due to low growth temperatures as well as difficulty in controlling the QD size uniformity and density[41].

When aiming to develop suitable systems for quantum light emitters, QDs based on local droplet etching (LDE) have been stated to be the predestined growth method[42]. This statement is grounded on the successes of LDE-based GaAs/AlGaAs QD emitters, which have consistently exhibited transform-limited exciton linewidths[43,44], exceedingly small inhomogeneous broadening thanks to QD size uniformity[45], bright single-photon emission[46], vanishing FSS[35], and highly indistinguishable and strongly entangled photons[7]. However, the bandgap energy of GaAs/AlGaAs

QDs naturally restricts their emission wavelength to around 800 nm. A recent report has demonstrated wavelength-tuning of LDE-based InGaAs/AlGaAs QDs up to ~900 nm through the addition of Indium to the QD material composition[47]. However, this is still clearly insufficient for reaching the telecom C-band wavelengths necessary for enabling efficient long-distance transfer of photonic qubit states. To solve this limitation, we have recently expanded the LDE growth method to fabricate (In)GaSb/AlGaSb QDs and demonstrated that they can emit at around 1.5 µm[48]. Moreover, we have proved that many beneficial properties associated with the LDE process can be preserved for (In)GaSb/AlGaSb LDE QDs. These include: (i) narrow linewidth excitonic emission[49], (ii) extremely low inhomogeneous broadening of ensemble QD emission spectra at 6-8 meV[48,49], (iii) single-photon emission capability with $g^{(2)}(0) = 0.16$ at 1470 nm for pure GaSb/AlGaSb[50] and $g^{(2)}(0) = 0.05 \pm 0.03$ at 1500 nm for InGaSb/AlGaSb[48] QDs. Furthermore, single-photon emission at 1515 nm was recently reported for GaSb-based LDE QDs[51]. Recent work on InGaAs/InP QDs presents similar benefits of adopting the LDE-based fabrication approach. However, the development of this QD system is in a relatively nascent growth optimization phase, with limited insight available into its optical properties[26,52].

Despite the well-developed field of GaAs/AlGaAs LDE QDs and the yet emerging LDE QD platforms around these other material systems, there is still a clear lack of understanding in the fundamental interrelations underpinning the emission properties of such QDs. Namely, only a preliminary understanding exists in terms of the relation between the QD size effects and emission wavelength, partially due to the difficulty of determining the true geometry of the QDs[49,53]. Moreover, the interplay between the QD material composition and emission properties is even less understood[47].

In this study, a quantitative and theoretically predictable relation between the emission wavelength of InGaSb/AlGaSb QDs and their composition and size is established. It is shown that telecom-band emission can be engineered across telecom S-C bands without sacrificing QD ensemble uniformity or single-dot quality. To this end, experimental photoluminescence (PL) characteristics and numerical modeling of electronic structure and optical properties are compared for a realistic QD geometry. Ensemble and single-QD emission properties are also studied with the focus being on QD uniformity and FSS, respectively. By varying the QD composition and QD size independently, the ability to control the emission wavelength across 1.48 μm to 1.55 μm region (telecom S-band to C-band) is demonstrated. This highlights a versatile engineering opportunity exploiting QD morphology and material for achieving desired emission properties at a target wavelength. In addition, we show that an excellent ensemble inhomogeneous broadening (< 7 meV) is maintained across the whole tuning range. We also prove that the emission characteristics of single-QDs show no clear evidence of degradation nor increase in FSS with In-composition, which is consistent with earlier findings for InGaAs/AlGaAs[47].

## MATERIALS AND METHODS

### *SAMPLE FABRICATION*

The samples were grown with solid-source molecular beam epitaxy on Te-doped GaSb(001) substrates. A schematic of the general sample structure is shown in Figure 1(a). The sample consists of a 100 nm GaSb buffer layer grown at 485 °C with a growth rate of 0.7 μm $h^{-1}$, followed by 200 nm lattice-matched $AlAs_{0.08}Sb_{0.92}$ barrier layer grown at 500 °C with a growth rate of 0.3 μm $h^{-1}$. The $AlAs_{0.08}Sb_{0.92}$ barrier prevents optically excited carriers from reaching the substrate, thus suppressing background luminescence originating from the GaSb substrate material. Subsequently, 100 nm of $Al_{0.3}Ga_{0.7}Sb$ is grown with a growth rate of 1 μm $h^{-1}$; during this step the temperature is ramped down from 500°C to 395°C to trigger formation of low-density single-QD sites determined by depositing Al droplets. The LDE process follows the steps reported in our earlier work[49], with the main step being the deposition of Al droplets under a small Sb-flux to etch the underlying $Al_{0.3}Ga_{0.7}Sb$ material. The formed nanoholes are then filled by deposition of an $In_xGa_{1-x}Sb$ layer, which also forms a quantum well (QW) on the planar areas between the nanoholes. The structure is then further covered by a 100 nm $Al_{0.3}Ga_{0.7}Sb$ layer at 395°C. In addition, for selected samples, the etching and/or filling process was repeated on top of the upper $Al_{0.3}Ga_{0.7}Sb$ layer to aid in the determination of the raw unfilled nanoholes and/or filled nanoholes dimensions and densities using atomic force microscope (AFM) and/or scanning electron microscopy (SEM).

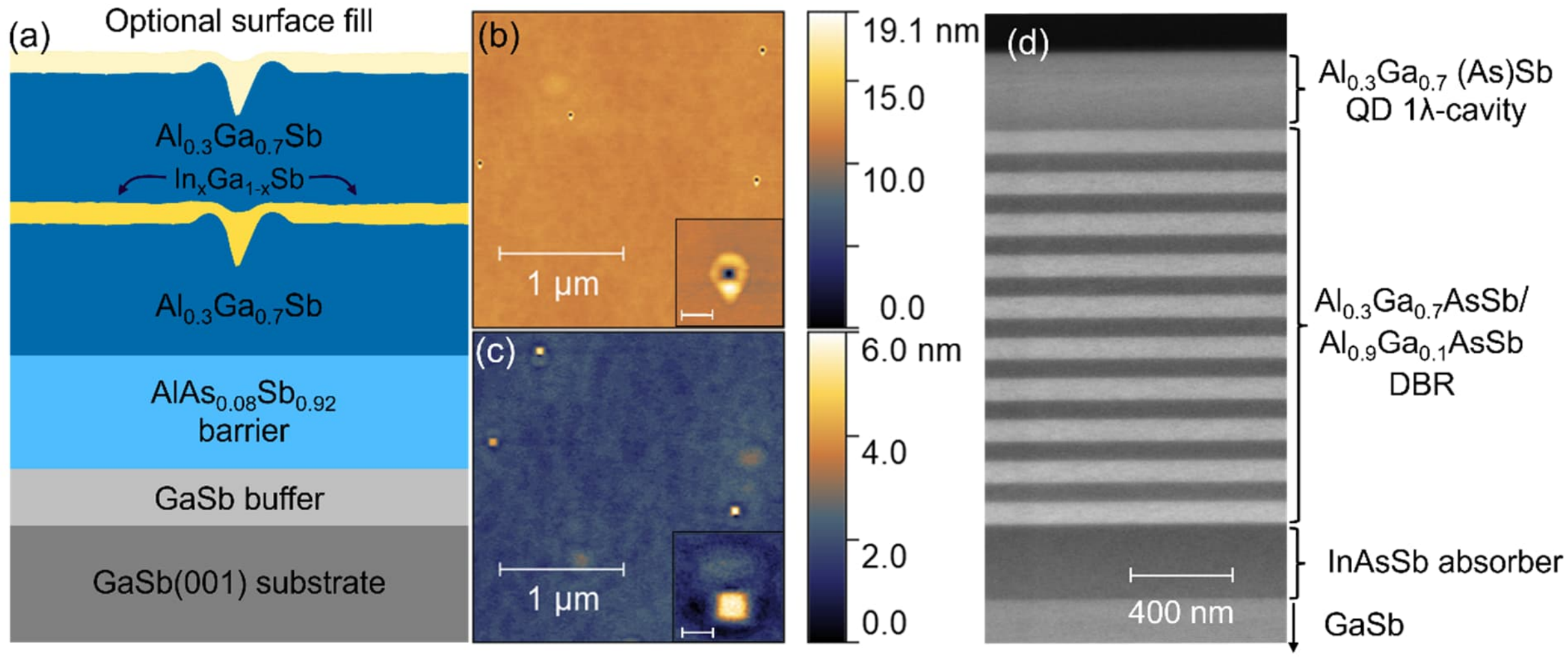

**Figure 1**. (a) Schematic illustration of the sample structure. (b-c) AFM images of the etched unfilled nanoholes (b) and 18 ML GaSb surface-filled nanoholes (c). The insets in the bottom right of figures (b-c) show close-up scans, with a scale bar of 50 nm. (d) Cross-sectional SEM image of the QD-cavity structure.

For all samples, the LDE process was kept identical to provide a consistent base of nanohole geometry and density for fabricating the $In_xGa_{1-x}Sb/Al_{0.3}Ga_{0.7}Sb$ QDs. The resulted nanoholes morphology as shown in Figure 1(b), revealing a diameter of about 50 nm, measured from the nanoring apex-to-apex distance in the [0-1-1] direction.

Two sample sets were grown to study the effects of wavelength tuning. The first sample set kept the filling layer thickness constant at 18.0 monolayers (ML) while varying the composition of the $In_xGa_{1-x}Sb$ filling layer, with x = 0.000, x = 0.039, x = 0.067, and x = 0.079. The In-content was tuned by varying the In-flux while keeping all other growth parameters constant. To promote preferential material diffusion into the nanoholes instead of the larger planar areas, the $In_xGa_{1-x}Sb$ layer was deposited in a cyclic fashion. More specifically, a cycle included deposition of around 1.8 ML of $In_xGa_{1-x}Sb$ followed by 10-second growth interruption to promote adatom diffusion. Thus, 10 cycles were repeated for a filling of 18.0 ML. The second sample set consisted of keeping the In-content at a constant value (x = 0.068) and varying the thickness of the $In_xGa_{1-x}Sb$ filling layer simply by adjusting the number of filling cycles. Three samples were grown with filling layer thicknesses of 10.8 ML, 14.4 ML, and 18.0 ML. We note that due to the limited material diffusion into the nanoholes, they were not completely planarized even with 18 ML deposition, as shown in Figure 1(c).

Additional samples were grown with a bottom distributed Bragg reflector (DBR) stack to improve the photon collection efficiency for single-QD spectroscopy. As illustrated in Figure 1(d), the QD layer is incorporated into an $Al_{0.3}Ga_{0.7}(As)Sb$ $\lambda$-cavity with a 9.5-pair $Al_{0.3}Ga_{0.7}AsSb/Al_{0.9}Ga_{0.1}AsSb$ bottom DBR. For the DBR sample structures an additional $InAs_{0.91}Sb_{0.09}$ layer with a thickness of 400 nm was deposited underneath the DBR as an absorber material to reduce the background PL signal originating from the n-GaSb substrate, which can overlap with the QD emission.

## OPTICAL CHARACTERIZATION

Photoluminescence spectra of the QD ensemble were measured in a closed-loop He-cooled cryostat at a temperature of ~12 K. The laser beam was targeted into a defocused spot diameter of 1.3 mm (full width at half-maximum; FWHM) on the sample surface. In this scenario, the laser spot is expected to simultaneously excite roughly $2.6\times10^5$ QDs (the QD density was determined using SEM; see Supplementary Information). PL emission was collected by a 0.5 m monochromator and detected with an InGaAs photodiode.

Single-QD spectroscopy was carried out with a high-resolution micro-PL (μ-PL) setup. The sample was mounted in a continuous-flow helium cryostat and cooled down to 10 K for low-temperature measurements. The QDs were excited using an external cavity tunable continuous-wave semiconductor laser filtered by a 15 cm focal length monochromator. The tuning range of this laser between 1425 – 1530 nm allows for quasi-resonantly exciting carriers into the QD excited states. The laser excitation was introduced into the main optical axis with a 50:50 non-polarizing beam splitter. The laser beam was focused on the sample surface into a spot of 1 μm diameter using a 20× magnification microscope objective characterized with a numerical aperture of 0.4 and a working distance of 20 mm. PL emission from the QD(s) was collected using the same microscope objective and guided to the detection system. The detection was provided by a 1 m focal length monochromator combined with a liquid nitrogen-cooled multichannel linear array InGaAs detector. The spectral resolution of the setup is 20 μeV.

## THEORETICAL SIMULATIONS

The structural data from AFM scans of representative empty and filled nanohole surfaces presented in Figure 2(a) were used for defining the physical shape of the QD for the modeling. A 230 × 230 × 320 computational mesh was used with spacings of 0.6 nm, 0.6, and 0.3 nm along the [100], [010] and [001] directions, respectively. First, a uniform QD material ($In_xGa_{1-x}Sb$ with varying In content $x$) was assumed between the two surfaces, and the barrier material $Al_{0.3}Ga_{0.7}Sb$ outside. Next, Gaussian averaging (with varying isotropic spatial extent corresponding to standard deviation $\sigma$) was used to simulate atom diffusion at interfaces. Even though the lattice constant mismatch between QD and bulk materials is relatively small, the strain was calculated by minimizing the elastic energy in the continuum elasticity theory. The strain induced by the GaSb substrate on which the sample has been grown was implemented using boundary conditions for the strain tensor at the bottom of the computational domain. Figure 2(a) shows an exemplary cross-section of the simulated QD, while Figures 2(b) and 2(c) show the calculated profiles of the volumetric and biaxial strain in the structure. Next, the piezoelectric field induced by the shear strain was calculated up to the second-order terms in the strain-tensor elements.

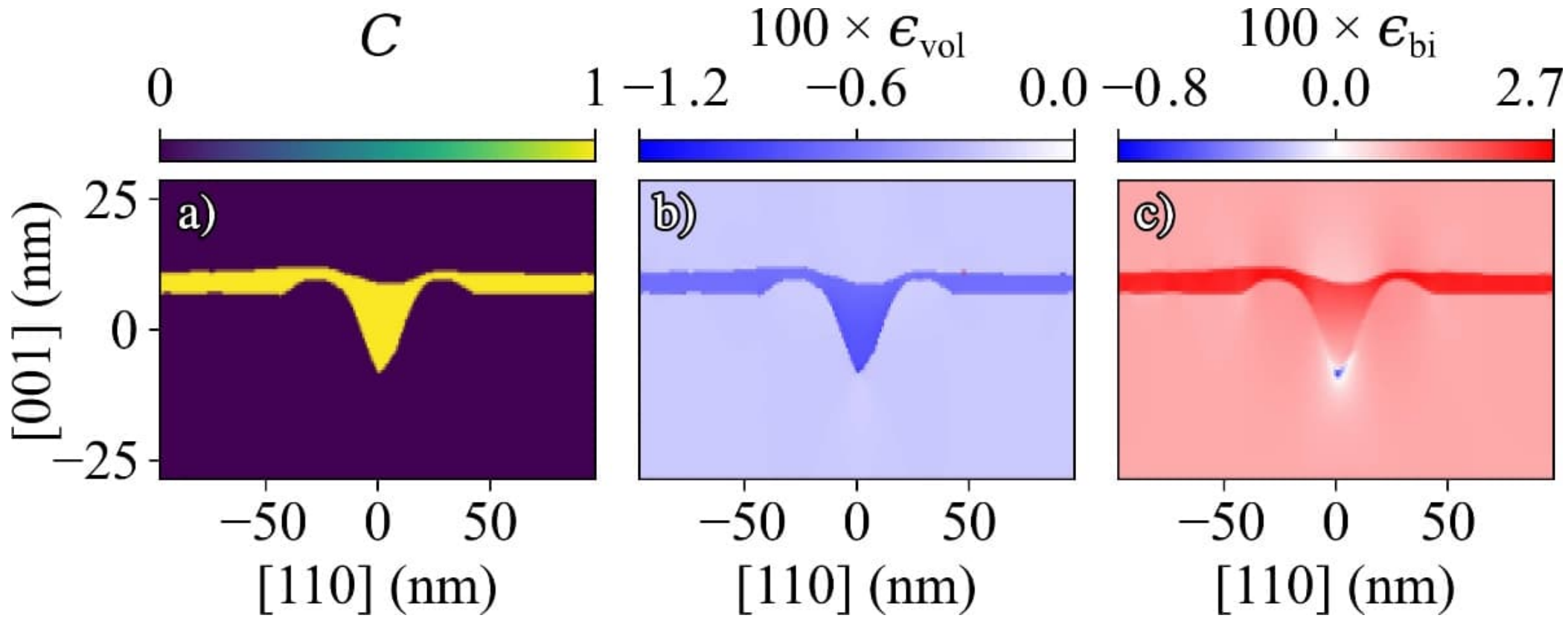

**Figure 2**. Cross-sectional view of a QD with 10.8 ML filling thickness, no material intermixing and QD indium concentration of $x$ = 0.1 in the (1-10) plane. a) Material composition. Value 0

represents $Al_{0.3}Ga_{0.7}Sb$, 1 represents $In_xGa_{1-x}Sb$. b) Volumetric strain $\varepsilon_{vol} = \varepsilon_{xx} + \varepsilon_{yy} + \varepsilon_{zz}$. c) Biaxial strain $\varepsilon_{bi} = -\varepsilon_{xx} - \varepsilon_{yy} + 2\varepsilon_{zz}$.

To calculate single-particle and carrier-complex states, we use a combined framework of multiband $\boldsymbol{k}\cdot\boldsymbol{p}$[54] and configuration interaction[55] methods in the envelope function approximation. For this, a custom computational code is used[56]. The complete form of the $\boldsymbol{k}\cdot\boldsymbol{p}$ Hamiltonian and the description of the numerical implementation can be found in reference[57]. The material parameters used, including their sources, are given in Table I and Table II, while the interpolation method used is described in reference[58]. The impact of structural strain is included via the Bir-Pikus Hamiltonian and quasi-momentum-dependent terms. The model also includes spin-orbit coupling. The states of interacting carriers (excitons) are calculated within the configuration interaction method using a 12×12 electron-hole configuration basis. This computation includes the direct Coulomb and phenomenological electron-hole exchange interactions.

*TABLE I. Material parameters of InSb, GaSb, and AlSb used in the calculations. Unless otherwise stated, parameters are from reference[59].*

| Parameter | InSb | GaSb | AlSb |
|---|---|---|---|
| $a$ (Å) | 6.46896 | 6.08174 | 6.1277 |
| $E_g$ (eV) | 0.235 | 0.812 | 2.386 |
| VBO (eV) | 0 | −0.03 | −0.41 |
| $m_e$ | 0.0135 | 0.039 | 0.14 |
| $E_P$ (eV) | 23.3 | 27.0 | 18.7 |
| Δ (eV) | 0.81 | 0.76 | 0.676 |
| $\gamma_1$ | 34.8 | 13.4 | 5.18 |
| $\gamma_2$ | 15.5 | 4.7 | 1.19 |
| $\gamma_3$ | 16.5 | 6.0 | 1.97 |
| $a_c$ (eV) | −6.94 | −7.5 | −4.5 |
| $a_v$ (eV) | −0.36 | −0.8 | −1.4 |
| $b_v$ (eV) | −2.0 | −2.0 | −1.35 |
| $d_v$ (eV) | −4.7 | −4.7 | −4.3 |
| $C_{11}$ ($10^2$ GPa) | 6.847 | 8.842 | 8.769 |
| $C_{12}$ ($10^2$ GPa) | 3.735 | 4.026 | 4.341 |
| $C_{44}$ ($10^2$ GPa) | 3.111 | 4.322 | 4.076 |
| $e_{14}$ (C/m$^2$) | −0.161[60] | −0.216[60] | −0.094[60] |
| $B_{114}$ (C/m2) | −0.62[60] | −0.31[60] | −0.76[60] |
| $B_{124}$ (C/m2) | −4.04[60] | −2.77[60] | −1.99[60] |
| $B_{156}$ (C/m2) | −0.16[60] | −0.70[60] | −0.82[60] |
| $\epsilon_r$ | 16.8[61] | 15.69[61] | 12.04[61] |
| $\epsilon_\infty$ | 15.68[61] | 14.44[61] | 10.24[61] |
| $n_r$ [$\lambda$ = 1.5 μm] | 3.9734[62] | 3.8389[62] | 3.2603[63] |

*TABLE II. Bowing parameters for ternary alloys from reference*[59]

| **Parameter** | **GaInSb** | **$Al_xGa_{1-x}Sb$** | **AlInSb** |
|---|---|---|---|
| $E_g$ (eV) | 0.415 | $-0.044 + 1.22x$ | 0.43 |
| $\Delta$ (eV) | 0.1 | 0.3 | 0.25 |
| $m_e$ | 0.0092 | 0 | 0 |

## RESULTS AND DISCUSSION

Figure 3(a) shows the ground state (GS) PL spectra of the $In_xGa_{1-x}Sb$ QD ensemble for the sample set with varying In-content (constant filling layer thickness). A monotonic redshift of the emission wavelength for the GS transition from 1482 nm towards 1550 nm with In content varying from $x$ = 0.000 to $x$ = 0.079 is observed. For all samples, the spectrum closely follows a Gaussian distribution [evident from solid red line fitting in Figure 3(a)], exhibiting low inhomogeneous broadening with FWHM of only around 5.7 ± 0.5 meV. This indicates that In-alloying is likely homogeneous throughout the QD population, and that QD uniformity is limited by the uniformity of the original nanohole population, which is kept identical for the samples. The inset in Figure 3(a) shows the determined shift in GS energy referenced to the In-free ($x$ = 0.0) sample, with a linear decrease in energy of −4.4 ± 0.3 meV/In%. For comparison, experimental data on InGaAs/AlGaAs LDE-QDs[47] show an emission energy shift of around −3.7 meV/In%, indicating a similar range of shifting behavior despite different QD morphology and materials.

Figure 3(b) shows the theoretical results obtained in a full calculation for excitons in QDs with different $\sigma$ as a function of the average In-content in the QD. In the simulation we assume that the average In-content in the QD results from an interplay of the material diffusion and changes in the nominal composition of the deposited material. It is calculated by averaging over the QD volume bounded by the nanohole and filling geometry. Apart from the expected decrease in the emission energy with increasing In-content, we find that material diffusion also has a direct effect. Namely, strong material intermixing increases the exciton energy as well as its sensitivity to changes in the average composition. A softer QD interface effectively makes the confinement narrower and shallower in the region with the largest carrier wave function weight, leading to a blueshift, as also found in GaSb-based quantum wells[64,65]. The increased sensitivity can be in turn traced back to the relatively weak carrier spatial confinement, causing carriers to be more sensitive to the exact shape of the bottom of the confinement influenced by the intermixing. The resulting theoretical shift of −4.3 ± 0.3 meV/In% with $\sigma = 0.6$ nm most closely matches the experimental value of −4.4 ± 0.3 meV/In%. This low value of intermixing is consistent with the generally low In-content used, far away from the immiscibility region[66], as well as the low growth temperature (395°C) used in the $Al_{0.3}Ga_{0.7}Sb$ QD capping layer, limiting annealing-induced intermixing. The likely explanation for the small discrepancy between the theoretical and experimental shifts could be that the In-content in the QD is enriched compared to the planar regions, given the high adatom mobility of In[67,68].

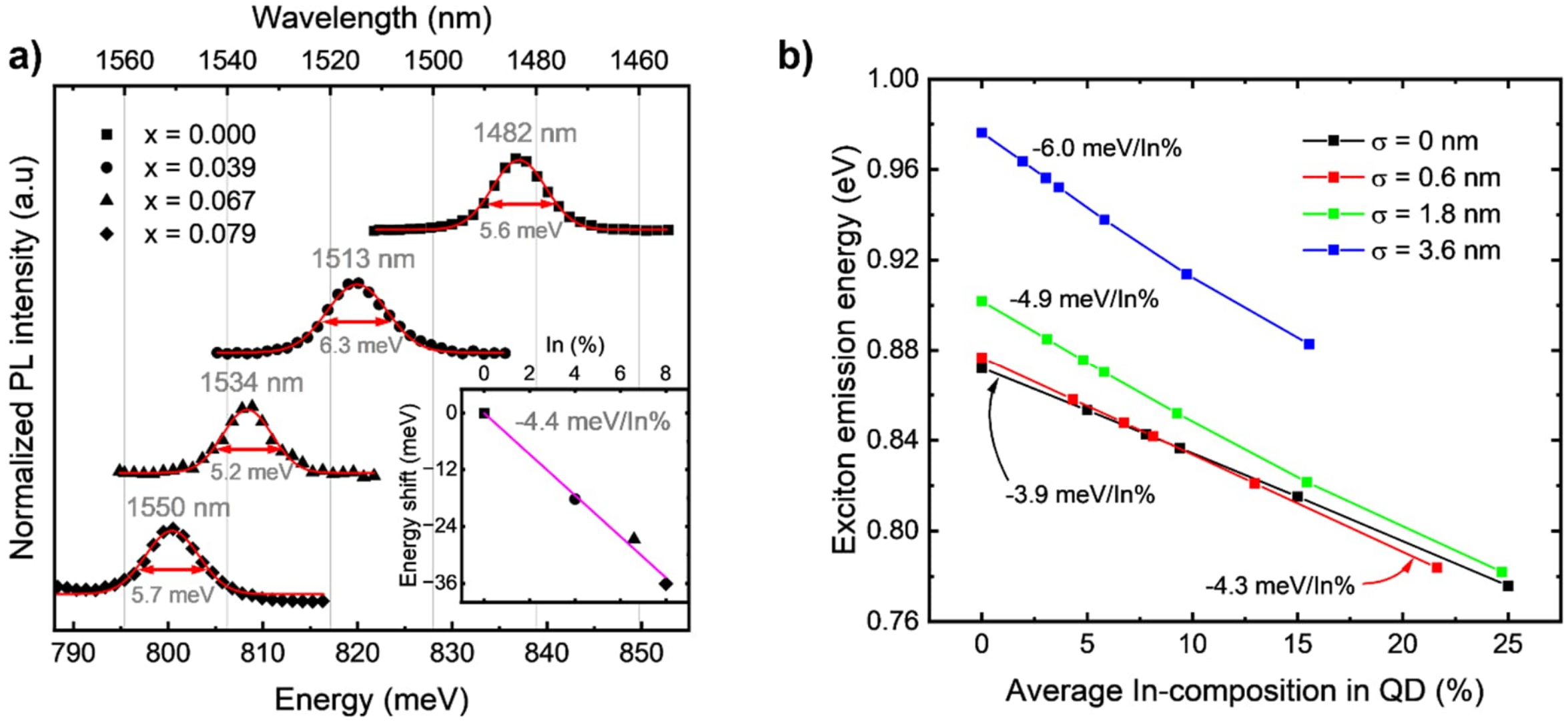

**Figure 3.** a) Ensemble PL emission spectra from $In_xGa_{1-x}Sb$ QDs with varying In-content and constant $In_xGa_{1-x}Sb$ filling thickness of 18 ML. The inset shows the determined GS energy shift (wrt. the In-free $x$ = 0.000 sample) as a function of In-content with a linear fit with a slope of −4.4 ± 0.3 meV/In%. b) Exciton emission energy as a function of average In-content in QD material with varying Gaussian averaging spatial extent $\sigma$.

Figure 4(a) shows the ensemble QD GS emission spectra for the second sample set, where the In-content was kept constant ($x$ = 0.068) while varying the thickness of the $In_xGa_{1-x}Sb$ nanohole filling layer. A monotonic redshift of the emission wavelength from 1521 nm to 1548 nm is observed with increase in filling layer thickness from 10.8 ML to 18.0 ML. The average linewidth of the GS emission peaks is 6.0 ± 0.1 meV for this sample set with minimal difference between the samples. This further indicates that the uniformity is likely not linked to the size or composition of the QDs but mainly controlled by the original nanohole dimensions, which serve as the template setting the uniformity of the QD population. We also note that generally all of the samples here exhibit narrow

GS ensemble emission, significantly narrower than those typically reported for SK QDs[69], GaAs/AlGaAs QDs[53], and GaSb/AlGaSb QDs[49]. The improvement in uniformity is a result of a more optimal growth procedure compared to our earlier work[49].

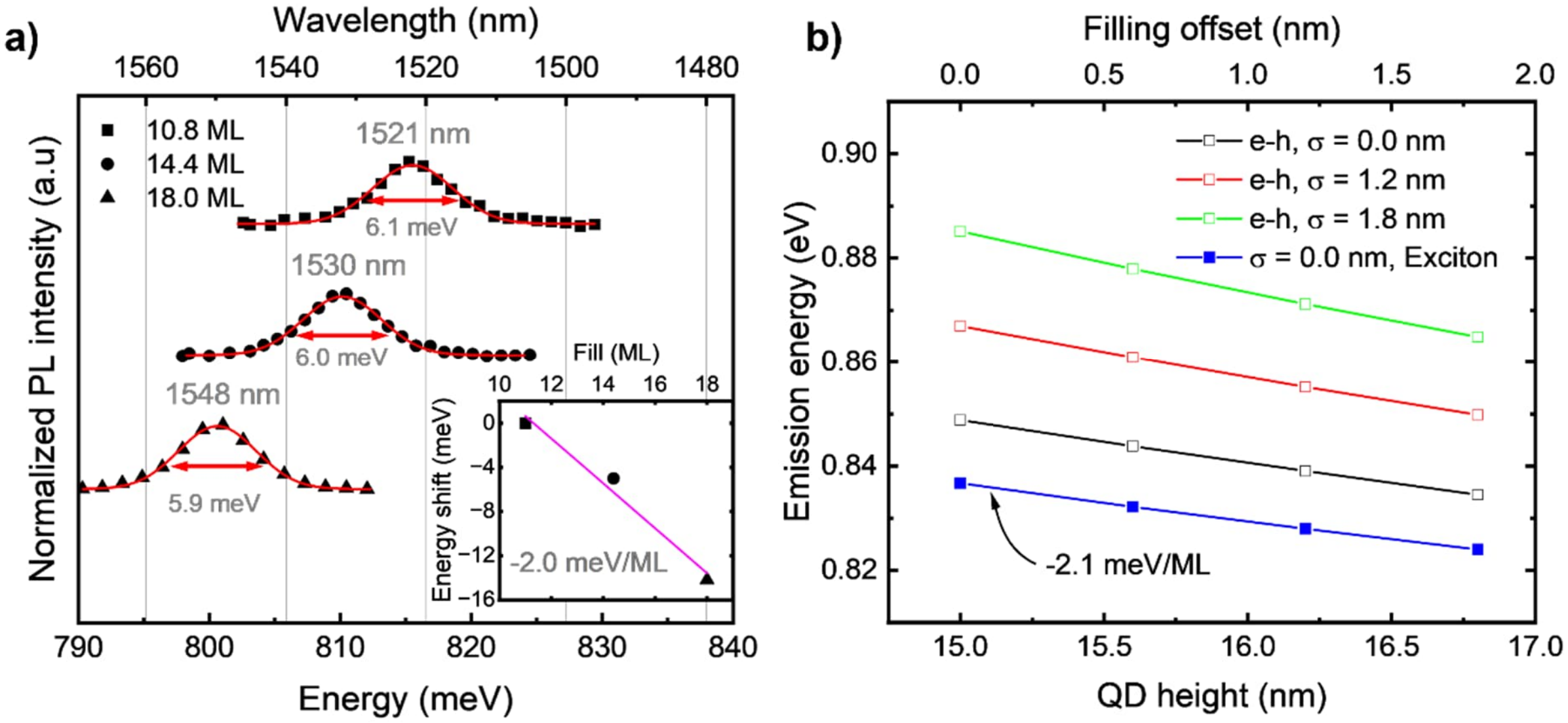

**Figure 4**. a) Ensemble PL emission spectra from $In_{0.068}GaSb$ QDs as a function of nanohole filling layer thickness. The inset shows the GS energy shift as a function of fill in ML with a slope of −2.0 ± 0.3 meV/ML. b) Calculated QD emission energy as a function of the additional deposited QD material ($In_{0.094}Ga_{0.906}Sb$) thickness (top axis) or QD height (bottom axis) for different Gaussian averaging spatial extents $\sigma$. Results labeled 'e-h' are estimated based on the electron and hole GS energy difference, while the 'exciton' series includes carrier interactions. The slope refers to the exciton series.

The GS energy shift with respect to the lowest filling layer thickness (10.8 ML) is shown in the inset of Figure 4(a). Given the limited amount of samples, we approximate the trend with a simple linear function with a determined slope of −2.0 ± 0.3 meV/ML, despite expectation of a non-linear shift based on a simple particle-in-a-box model. This allows us to translate our 6 meV ensemble emission FWHM into ~1.5 ML QD height uncertainty. In reality, there are likely nonlinear factors present from preferential filling of the nanoholes due to the cyclic growth sequence used in aiding adatom diffusion, causing a compounding effect in the true size of the QD. We note that GaAs/AlGaAs LDE-QDs have shown a slightly non-linear relation of QD emission energy as a function of QD height[53].

We simulate this variation in QD height, obtained in the growth by depositing different amounts of QD material, in our model. For this, we assume that changing the nominal amount of deposited material corresponds to a uniform vertical shift of the top QD surface such that the thickness of the 2D layer far from QD equals the nominal deposition thickness in the actual QD growth. Figure 4(b) shows the results for three different diffusion spatial extents (hollow points labeled ’e-h’). Based on the best agreement with the experimentally observed emission energies, we use the case with approximate lack of diffusion for a subsequent full calculation of the excitonic energy (filled points labeled ’exciton’). The results show linear reduction in the exciton emission energy with QD height due to the weakened carrier spatial confinement. The spatial extent of material diffusion weakly affects this trend, causing a parallel shift of the dependence. The resulting excitonic calculation without diffusion, with a slope of −2.1 meV/ML, very closely reproduces the experimentally measured slope of −2.0 ± 0.3 meV/ML.

Figure 5 shows PL emission spectra of single QDs in the sample set grown with varying In-contents ($x = 0.000$, $x = 0.039$, $x = 0.066$, $x = 0.081$) and constant filling thickness of 18.0 ML. Single QDs in all samples exhibit a few distinct narrow excitonic emission lines together with an ensemble of charged and neutral multi-excitonic lines, which are also enhanced by the cavity and thus produce a central envelope around the excitonic lines. Notably, the envelope of excitonic emission lines red-shift from 1479 nm to around 1547 nm as a function of increasing In-content.

This trend closely follows that of the shift observed for the ensemble GS energy in Figure 3(a). At the same time, no clear trend in broadening of the sharp excitonic emission lines is observed.

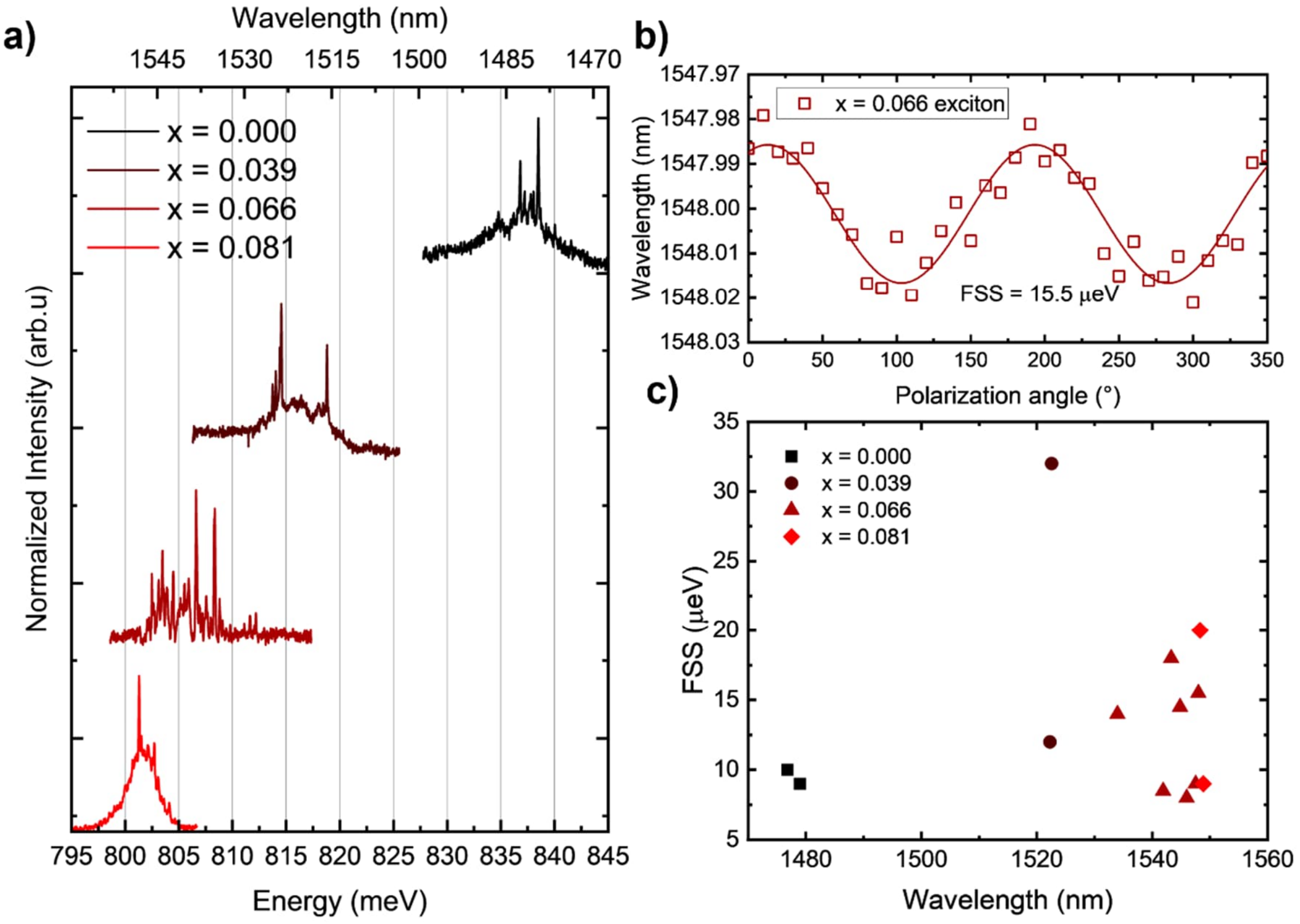

**Figure 5**. a) Single-QD PL of quasi-resonantly pumped $In_xGa_{1-x}Sb$ QDs within the DBR-structures with varying In-contents and constant filling thickness of 18.0 ML. Background substraction was done for sample x = 0.081 as the emission started to overlap with the background emission of the GaSb substrate. b) Example FSS determined from a sinusoidal fit with a center wavelength of 1548 nm for a QD with an In-content $x$ = 0.066. c) FSS as a function of exciton emission wavelength obtained from QD-cavity samples with In-content of $x$ = 0.000, 0.039, 0.066, 0.081 with fixed $In_xGa_{1-x}Sb$ filling thickness of 18.0 ML.

At least two locations with distinct excitonic emission for each QD sample were measured in polarization-resolved µ-PL to determine their FSS values. The collected FSS values for the samples are shown in Figure 5(c) as a function of the wavelength of each specific excitonic line. Again, for each sample, the excitonic lines are grouped closely together, reflecting the uniformity of the QDs. In terms of FSS, while there is clear QD-to-QD variation for each sample, there is no clear overall trend that would indicate an impact of In-content on the average FSS value. Namely, the QD samples with higher In-content can reach the same FSS values of <10 µeV as with the In-free GaSb QDs. Furthermore, the average FSS value is around 13.8 µeV with a standard deviation of 6.7 µeV, and a minimum of 8.0 µeV. It is worth mentioning that a spectrometer and not interferometric methods are used to measure the FSS, which limits the spectral resolution. Moreover, we have not used coherent excitation scheme which would improve the FWHM of emission lines, therefore we were not able to resolve very low FSS values (close to zero). The FSS values obtained fit well to our previous results on InGaSb/AlGaSb QDs[48] as well as the FSS values reported for InGaAs/AlGaAs QDs[47], yet fall short of the lowest FSS values reported for GaAs/AlGaAs QDs[8,35]. The higher FSS values here are likely related to the asymmetry of the nanohole geometry observable in Figure 1(b), which can be made symmetric through growth optimization[35]. For polarization-entangled photon sources, the generally considered pathway for reaching vanishing FSS is achieving sufficiently low FSS <10 µeV, as obtained here, and then tuning it to zero with external factors, e.g., by piezo-induced strain[70].

In conclusion, we have shown theoretically predictable effects of In-content and QD size on the emission wavelength of LDE-grown $In_xGa_{1-x}Sb$/AlGaSb QDs. When increasing the In-content experimentally, a linear shift in the GS energy of the QD of −4.4 ± 0.3 meV/In% is observed. This magnitude of shift closely matches the simulated value of −4.3 meV/In%. In terms of QD size, experiments showed GS emission energy shift of −2.0 ± 0.3 meV/ML when increasing the degree of nanohole filling, which matched the simulated value of −2.1 meV/ML. We experimentally observed consistent low inhomogeneous broadening of the ensemble QD emission, well below 7 meV throughout the demonstrated tuning range from 1482 nm to 1550 nm. The uniformity is an essential feature enabling reproducibility and scalability of quantum devices as well as two-photon interference from remote sources, which is an indispensable resource for quantum repeater architectures. Finally, we investigated single-QD emission properties as a function of QD In-content, which showed narrow excitonic emission lines with low FSS values of 13.8±6.7 µeV. Our results offer a framework for tuning and optimizing the exact morphology of $In_xGa_{1-x}Sb$/AlGaSb LDE-QDs for application-specific purposes in the telecom S and C bands.

ACKNOWLEDGEMENT

Z. O., T. P. and M. G. are grateful to Krzysztof Gawarecki for sharing his computational code. The collaborative work was carried out within the “FiGAnti” project, which is part of the QuantERA II EU Program (Grant Agreement No. 101017733) with support from the Strategic Research Council of Finland (Decision No. 361293) and National Science Centre Poland (Project 2023/05/Y/ST3/00125). Tampere University team also acknowledges the support received from the Strategic Research Council of Finland via “CryoLight” project (Decision No. 357351) and Flagship Program PREIN (Decision No. 368650) and Business Finland via TeleQuant Rise to Challenge project (Decision No. 1835/31/2025). Calculations have been carried out using resources provided by the Wroclaw Centre for Networking and Supercomputing (https://wcss.pl).